\documentclass[pre,preprint,showpacs,a4paper]{revtex4}
\usepackage{graphicx}

\begin{document}
\title{Optically bound microscopic particles in one dimension}
\author{ D. McGloin$^{1*}$, A.E. Carruthers$^1$, K. Dholakia$^1$ and E.M. Wright$^{1,2}$,}

 \affiliation{$^1$School of Physics and
Astronomy, University of St. Andrews, North Haugh, St. Andrews,
KY16 9SS, UK}
 \affiliation{$^2$Optical Science Center, University of Arizona, Tucson, AZ 85721, USA}

\date{\today}
%
\begin{abstract}
Counter-propagating light fields have the ability to create
self-organized one-dimensional optically bound arrays of
microscopic particles, where the light fields adapt to the
particle locations and vice versa. We develop a theoretical model
to describe this situation and show good agreement with recent
experimental data (Phys. Rev. Lett. 89, 128301 (2002)) for two and
three particles, if the scattering force is assumed to dominate
the axial trapping of the particles. The extension of these ideas
to two and three dimensional optically bound states is also
discussed.
\end{abstract}

\pacs{82.70.-y,45.50.-j, 42.60.Jf, 87.80.Cc} \maketitle

\section{Introduction}
The ability of light to influence the kinetic motion of
microscopic and atomic matter has had a profound impact in the
last three decades. The optical manipulation of matter was first
seriously studied by Ashkin and co-workers in the 1970s
\cite{Ashkin1,Ashkin2,Ashkin3}, and led ultimately to the
demonstration of the single beam gradient force trap
\cite{Ashkin4}, referred to as optical tweezers, where the
gradient of an optical field can induce dielectric particles of
higher refractive index than their surrounding medium to be
trapped in three dimensions in the light field maxima
\cite{Ashkin4}. Much of Ashkin's early work centered not on
gradient forces, but on the use of radiation pressure to trap
particles \cite{Ashkin1}, and a dual beam radiation pressure trap
was demonstrated in a which single particle was confined. This
work ultimately contributed to the development of the
magneto-optical trap for neutral atoms \cite{Chu}.

Recently we observed one-dimensional {\it arrays} of silica
spheres trapped in a dual beam radiation pressure trap
\cite{Sveta}. These arrays had an unusual property in that the
particles that formed the array were regularly spaced from each
other. The particles were redistributing the incident light field,
which in turn redistributed the particle spacings, allowing them
to reside in equilibrium positions. This effect, known as
\textquotedblleft optically bound matter" was first realised in a
slightly different context via a different mechanism to ours some
years ago \cite{Burns,Burns1} using a single laser beam and was
explained as the interaction of the coherently induced dipole
moments of microscopic spheres in an optical field creating bound
matter.

In the context of our study optically bound matter is of interest
as it relates to the way in which particles interact with the
light field in extended optical lattices, which may prove useful
for the understanding of three-dimensional trapping of colloidal
particles \cite{vanB}. Indeed optically bound matter may provide
an attractive method for the creation of such lattices that are
not possible using interference patterns. Bound matter may also
serve as a test bed for studies of atomic or ionic analogues to
our microscopic system \cite{Blatt}.

Subsequent to our report a similar observation was  made in an
experiment making use of a dual beam fiber trap \cite{Singer}. In
this latter paper a theory was developed that examined particles
of approximately the same size as the laser wavelength involved.
In this paper we develop a numerical model that allows us to
simulate the equilibrium positions of two and three particles in a
counter-propagating beam geometry, where the particle sizes are
larger than the laser wavelength, and fall outside the upper bound
of the limits discussed in \cite{Singer}. The model can readily be
extended to look at larger arrays of systems. We discuss the role
of the scattering and refraction of light in the creation of
arrays. In the next section we describe the numerical model we use
for our studies and derive predictions for the separation of two
and three spheres of various sizes. We then compare this with both
previous and current experiments.

\section{Theory Section}
\tolerance = 1000
 \maketitle
Our model comprises two monochromatic laser fields of frequency
$\omega$ counter-propagating along the z-axis which interact with
a system of $N$ transparent dielectric spheres of mass $m$,
refractive-index $n_s$, and radius $R$, with centers at positions
$\{\vec{r}_j(t)\}, j=1,2,\ldots N$, and which are immersed in a
host medium of refractive-index $n_h$. The electric field is
written
\begin{equation}
\vec{E}(\vec{r},t) = \frac{\hat{\bf e}}{2}\left[({\cal
E}_+(\vec{r})e^{ikz}+{\cal E}_-(\vec{r})e^{-ikz})e^{-i\omega t} +
c.c \right ] , \label{Efield}
\end{equation}
where $\hat{\bf e}$ is the unit polarization vector of the field,
${\cal E}_\pm(\vec{r})$ are the slowly varying electric field
amplitudes of the right or forward propagating $(+)$ and left or
backward propagating $(-)$ fields, and $k=n_h\omega/c$ is the
wavevector of the field in the host medium. The incident fields
are assumed to be collimated Gaussians at longitudinal coordinates
$z=-L/2$ for the forward field and $z=L/2$ for the backward field
\begin{equation}
{\cal E}_+(x,y,z=-L/2)={\cal
E}_-(x,y,z=L/2)=\sqrt{\frac{4P_0}{n_hc\epsilon_0\pi
w_0^2}}e^{-r^2/w_0^2} , \label{BCond}
\end{equation}
where $r^2=x^2+y^2$, $w_0$ is the initial Gaussian spot size, and
$P_0$ is the input power in each beam. It is assumed that all the
spheres are contained between the beam waists within the length
$L>>R$.

Consider first that the dielectric spheres are in a fixed
configuration at time $t$ specified by the centers
$\{\vec{r}_j(t)\}$. Then the dielectric spheres provide a
spatially inhomogeneous refractive index distribution which can be
written in the form
\begin{equation}
n^2(\vec{r}) = n_h^2 +
(n_s^2-n_h^2)\sum_{j=1}^{N}\theta(R-|\vec{r}-\vec{r}_j(t)|) ,
\label{RefInd}
\end{equation}
where $\theta(R-|\vec{r}-\vec{r}_j(t)|)$ is the Heaviside step
function which is unity within the sphere of radius $R$ centered
on $\vec{r}=\vec{r}_j(t)$, and zero outside, and $n_s$ is the
refractive-index of the spheres. Then, following standard
approaches \cite{Feit}, the counter-propagating fields evolve
according to the paraxial wave equations
\begin{equation}
\pm\frac{\partial{\cal E}_\pm}{\partial z} =
\frac{i}{2k}\nabla_\perp^2{\cal E}_\pm +
ik_0\frac{(n^2(\vec{r})-n_h^2)}{2n_h}{\cal E}_\pm , \label{Parax}
\end{equation}
along with the boundary conditions in Eq. (\ref{BCond}), where
$k_0=\omega/c$ and $\nabla_\perp^2=\partial^2/\partial
x^2+\partial^2/\partial y^2$ is the transverse Laplacian
describing beam diffraction. Thus, a given configuration of the
dielectric spheres modifies the fields ${\cal E}_\pm(\vec{r})$ in
a way that can be calculated from the above field equations. We
remark that even though the spheres move, and hence so does the
refractive-index distribution, the fields will always
adiabatically slave to the instantaneous sphere configuration.

To proceed we need equations of motion for how the sphere centers
$\{\vec{r}_j(t)\}$ move in reaction to the fields. The
time-averaged dipole interaction energy \cite{Ashkin4}, relative
to that for a homogeneous dielectric medium of refractive-index
$n_h$, between the counter-propagating fields and the system of
spheres is given by
\begin{eqnarray}
U(\vec{r}_1,\ldots,\vec{r}_N) &=& \int_{ }^{ }dV \epsilon_0\left
[n^2(\vec{r})-n_h^2 \right ]<\vec{E}^2> \nonumber \\ &=&
-\frac{\epsilon_0}{4}(n_s^2-n_h^2)\sum_{j=1}^{N}\int_{ }^{
}dV\theta(R-|\vec{r}-\vec{r}_j(t)|)\left[|{\cal
E}_+(\vec{r})|^2+|{\cal E}_-(\vec{r})|^2 \right ] , \label{U}
\end{eqnarray}
where the angled brackets signify a time-average which kills
fast-varying components at $2\omega$. The most important concept
is that the dipole interaction potential depends on the spatial
configuration of the spheres $U(\vec{r}_1,\ldots,\vec{r}_N)$ since
the counter-propagating fields themselves depends on the sphere
distribution via the paraxial wave equations (\ref{Parax}). This
form of the dipole interaction potential (\ref{U}) shows
explicitly that we pick up a contribution from each sphere
labelled $j$ via its interaction with the local intensity.
Assuming over-damped motion of the spheres in the host medium with
viscous damping coefficient $\gamma$, the equation of motion for
the sphere centers become
\begin{equation}
\gamma \frac{d\vec{r_j}}{dt}=\vec{F}_{grad,j} + \vec{F}_{scatt,j}
,\quad \vec{F}_{grad,j}=-\nabla_jU(\vec{r}_1,\ldots,\vec{r}_N)
\label{Newt}
\end{equation}
where $\nabla_j$ signifies a gradient with respect to $\vec{r}_j$,
and $\vec{F}_{grad,j},\vec{F}_{scatt,j}$ are the gradient and the
scattering forces experienced by the j$^{th}$ sphere, the latter
of which we shall give an expression for below.

Carrying through simulations for a 3D system with modelling of the
electromagnetic propagation in the presence of many spheres poses
a formidable challenge, so here we take advantage of the symmetry
of the system to reduce the calculation involved. First, for the
cylindrically symmetric Gaussian input beams used here we assume
that the combination of the dipole interaction potential, and
associated gradient force, and the scattering force supplies a
strong enough transverse confining potential that the sphere
motion remains directed along the z-axis. This means that the
positions of the sphere centers are located along the z-axis
$\vec{r}_j(t)=\hat{\bf z}z_j(t)$, and that the gradient and
scattering forces are also directed along the z-axis
$\vec{F}_j=\hat{\bf z}F_j$. Second, we assume that the sphere
distribution along the z-axis is symmetric around $z=0$, the beam
foci being at $z=\pm L/2$. This means, for example, that for one
sphere the center is located at $z=0$, for two spheres the centers
are located at $z=\pm D/2$, $D$ being the sphere separation
distance, and for three spheres the centers are at $z=0,\pm D$.
For three or less spheres the symmetric configuration of spheres
is captured by the sphere spacing $D$, and we shall consider this
case here. For more than three spheres the situation becomes more
complicated and we confine our discussion to the simplest cases of
two and three spheres.

With the above approximations in mind the equations of motion for
the sphere centers become
\begin{equation}
\gamma \frac{dz_j}{dt} = F_{grad,j}+ F_{scatt,j} , \qquad
j=1,2,\dots,N .
\end{equation}
At this point it is advantageous to consider the case of two
spheres, $N=2$, to illustrate how calculations are performed. For
a given distance $D$ between the spheres we calculate the
counter-propagating fields between $z=[0,L]$ using the beam
propagation method. From the fields we can numerically calculate
the dipole interaction energy $U(D)$ for a given sphere
separation, and the resulting axial (z-directed) gradient force is
then $F_{grad}(D)=-\partial U/\partial D$. Thus, by calculating
the counter-propagating fields for a variety of sphere separations
we can numerically calculate the gradient force which acts on the
relative coordinate of the two spheres. For our system we
approximate the scattering force \cite{Rohrbach} along the
positive z-axis for the j$^{th}$ sphere as
\begin{equation}
F_{scatt,j} \approx \left (\frac{n_h}{c} \right ) \left
(\frac{\sigma}{\pi R^2} \right ) \int_0^R 2\pi rdr
\frac{\epsilon_0 n_hc}{2}\left[|{\cal E}_+(x,y,z_j)|^2-|{\cal
E}_-(x,y,z_j)|^2 \right ] ,
\end{equation}
with $\sigma$ the scattering cross-section. This formula is
motivated by the generic relation $F_{scatt}=n_hP_{scatt}/c$ for
unidirectional propagation, with the scattered power
$P_{scatt}=\sigma I_0$, and $I_0$ the incident intensity. The
integral yields the difference in power between the two
counter-propagating beams integrated over the sphere
cross-section, and when this is divided by the sphere
cross-sectional area $\pi R^2$ we get the averaged intensity
difference over the spheres. For the case of two spheres we
calculate the scattering force $F_{scatt}(D)$, evaluated at the
position of the sphere at $z=D/2$, and for a variety of sphere
spacings $D$. A similar procedure can readily be applied to the
case of three spheres.

The theory described above has some limitations that we now
discuss. First, we assume that the spheres are trapped on-axis by
a combination of the scattering and/or dipole forces acting
transverse to the propagation axis. For this to be possible we
require that the sphere diameter be less than the laser beam
diameter $2w_0>D$. Furthermore, we have assumed paraxial
propagation that neglects any large angle or back-scattering of
the laser fields. However, when light is incident on a sphere of
diameter $D$ there is an associated wavevector uncertainty $\Delta
K D\simeq 2\pi$, and when $\Delta K\simeq 2k$ back-scattering can
occur, as it is within the uncertainty that an incident wave of
wavevector $k$ along a given direction is converted into $-k$.
This yields the condition $D>\lambda/2n_h$, with $\lambda$ the
free-space wavelength, to avoid back-scattering and so that our
paraxial assumptions are obeyed.

Our goal is to compare the axial gradient and scattering forces
for an array of two and three spheres and compare with the
experimental results. However, the scattering cross-section for
our spheres, which incorporates all sources of scattering in a
phenomenological manner, cannot be calculated with any certainty.
Our approach, therefore, will be to calculate the equilibrium
sphere separation $F(D)=0$ for the gradient and scattering forces
separately, which does not depend on the value of the
cross-section, and compare the calculated sphere separations with
the experimental values. By comparing the theoretical predictions
with the experiment for $N=2,3$ we can determine the dominant
source of the axial force acting on the spheres.

\section{Experiment}

To compare our theory with experiment we use data from our
previous work \cite{Sveta} and also recreate that experiment, but
using a different laser wavelength and particle sphere size. The
previously reported experiment \cite{Sveta} makes use of a
continuous-wave 780nm Ti:Sapphire laser, which is split into two
beams with approximately equal power (25mW) in each arms. Each of
the beams is focussed down to a spot with a 3.5 $\mu m$ beam waist
and then passed, counterpropagating, through a cuvette with
dimensions of 5mm x 5mm x 20mm. The beam waists were separated by
a finite amount, which is discussed further below. Uniform silica
spheres with a 3$\mu m$ diameter (Bangs Laboratories, Inc) in a
water solution were placed in the cuvette, and the interaction of
the beams with the sample caused one-dimensional arrays of
particle to be formed. The refractive index of the spheres is
approximately 1.43. We also carried out a similar experiment using
a 1064$nm$ Nd:YAG laser where the beam waists were 4.3$\mu m$ and
we used 2.3$\mu m$ diameter spheres. The particles were viewed by
looking at the scattered light orthogonal to the laser beam
propagation direction viewed on a CCD camera with an attached
microscope objective (x20, NA=0.4, Newport).

To compare our theory with experimental results we need to
concentrate on a small number of parameters, the sphere size, the
beam waist, the refractive index of the spheres and the beam waist
separation. We know the particle sizes and can make a good
estimate as to their refractive index, further we can measure the
beam waist to a high degree of accuracy. The only problematic
factor is the beam waist separation. Due to experimental
constraints, this is quite difficult to measure. We estimate the
waist separation by filling the cuvette with a high density
particle solution and looking at the scattered light from the
sample. The high density of particle allows us to map out the
intensity pattern of the two beams and hence make an estimate as
to the waist separation. This is, however, an inaccurate method
and leaves us with an error of more than 100\%. We therefore use
our model to help us fix the beam waist separation on a single
result and then examine the behavior of the model when varying
other parameters. The error in the beam waist separation is not as
extreme as first it sounds however. Modelling the system for a
range of beam waist separations from 80$\mu$m to 200$\mu$m results
in a predicted range of sphere separations as shown in figure 1,
2.3$\mu$m diameter spheres. We see that although initially the
beam waist separation difference makes a reasonable difference to
the predicted sphere separation the region that we believe we are
working in, $\sim180\mu$m waist separation, is relatively flat.
Therefore even if we do have a large error in this value, the
predicted result does not vary significantly. This increases our
confidence that we have the correct beam waist separation with a
higher uncertainty that our experimental measurements of this
parameter suggests.

We begin by examining the case of the 2.3 $\mu m$ diameter
spheres.

\begin{figure}
\centerline{\includegraphics[width=8cm]{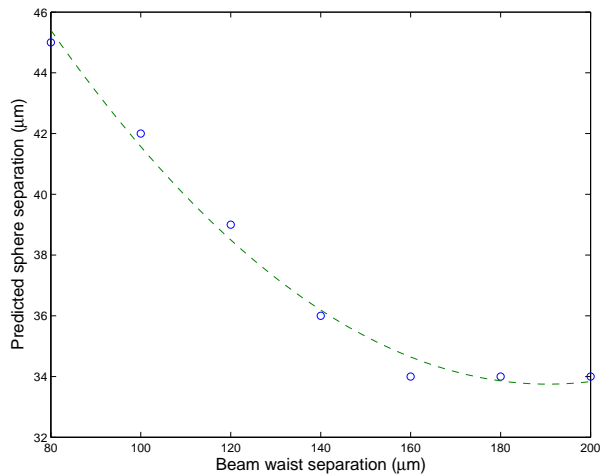}}
\caption{Sphere separation as a function of beam waist separation
for 2 two 2.3$\mu m$ spheres. The rate of change of sphere
separation is seen to drop off as the waist separation increases.
The fit to a parabola is to aid the eye, rather than to suggest a
quantitative relationship.}
\end{figure}

\subsection{2.3 micron micrometer diameter spheres}
We consider the case for chains of both two and three spheres. For
two spheres we measure a sphere separation of ~34$\mu m$, for a
beam waist, $\omega_0=4.3\mu m$ at a laser wavelength,
$\lambda=1064nm$. Using a beam waist separation of $180\mu m$ our
model predicts an equilibrium in the scattering force of 34$\mu
m$, as is shown in figure 2. The intensity in the x-z plane for
this configuration is shown in figure 3. We see no such
equilibrium in the gradient force, shown in figure 4 and conclude
that the scattering force is the dominant factor in this instance.
Using the same parameters for the three sphere case give us a
sphere separation prediction of ~62$\mu m$, as shown in figure 5.
Again this dominates over the gradient force, this assumption
being valid, as the theory gives a good prediction of our
experimental observations. Our experimental result is 57$\mu m$,
but we estimate our model value falls within the standard
deviation error we observe on our experimental measurements.

\begin{figure}
\centerline{\includegraphics[width=8cm]{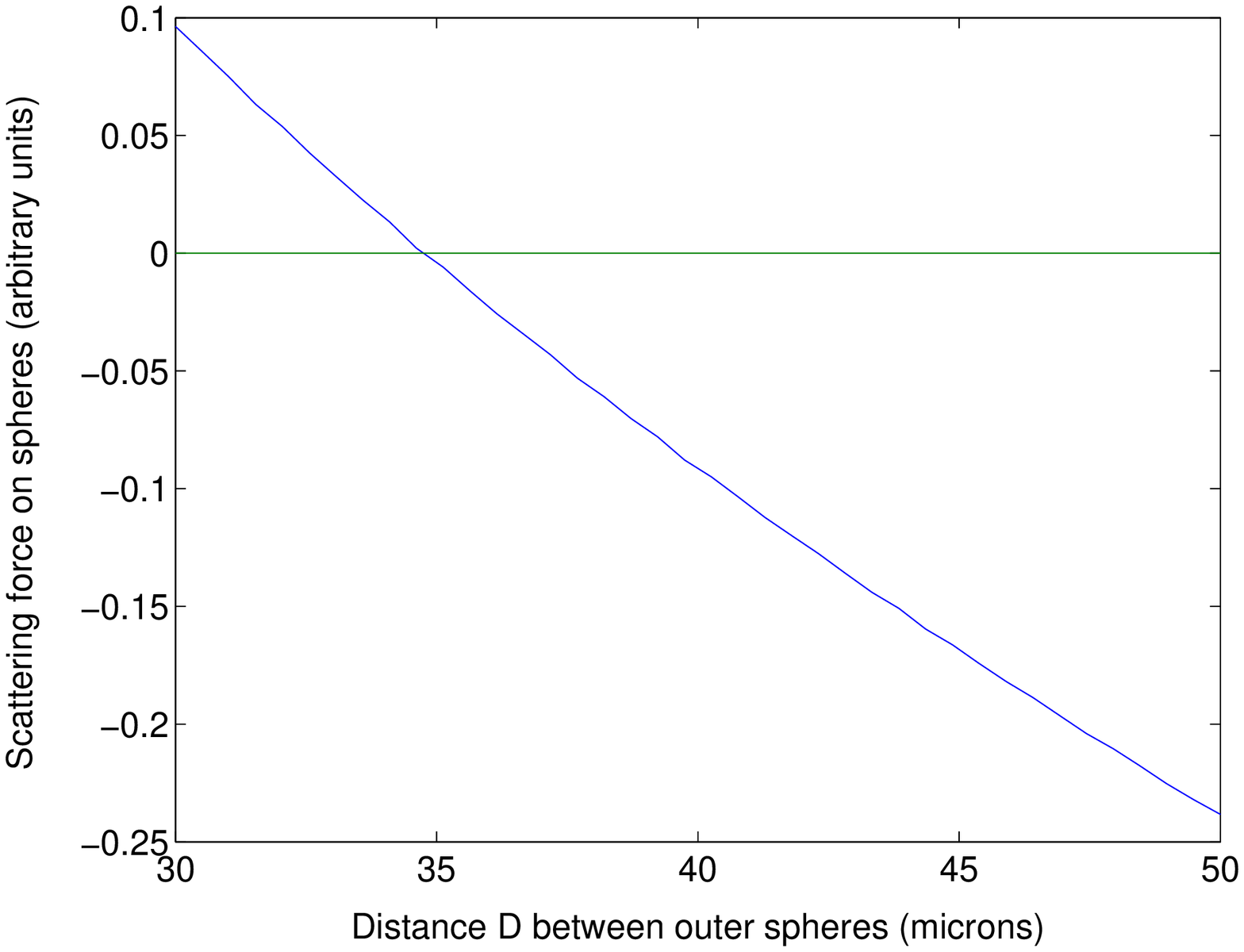}}
\caption{Scattering force on two 2.3 micron $\mu m$ diameter
silica spheres with the beam waists 180 $\mu m$ apart.
$\omega_0=3.5\mu m$ and $\lambda=1064nm$.}
\end{figure}

\begin{figure}
\centerline{\includegraphics[width=8cm]{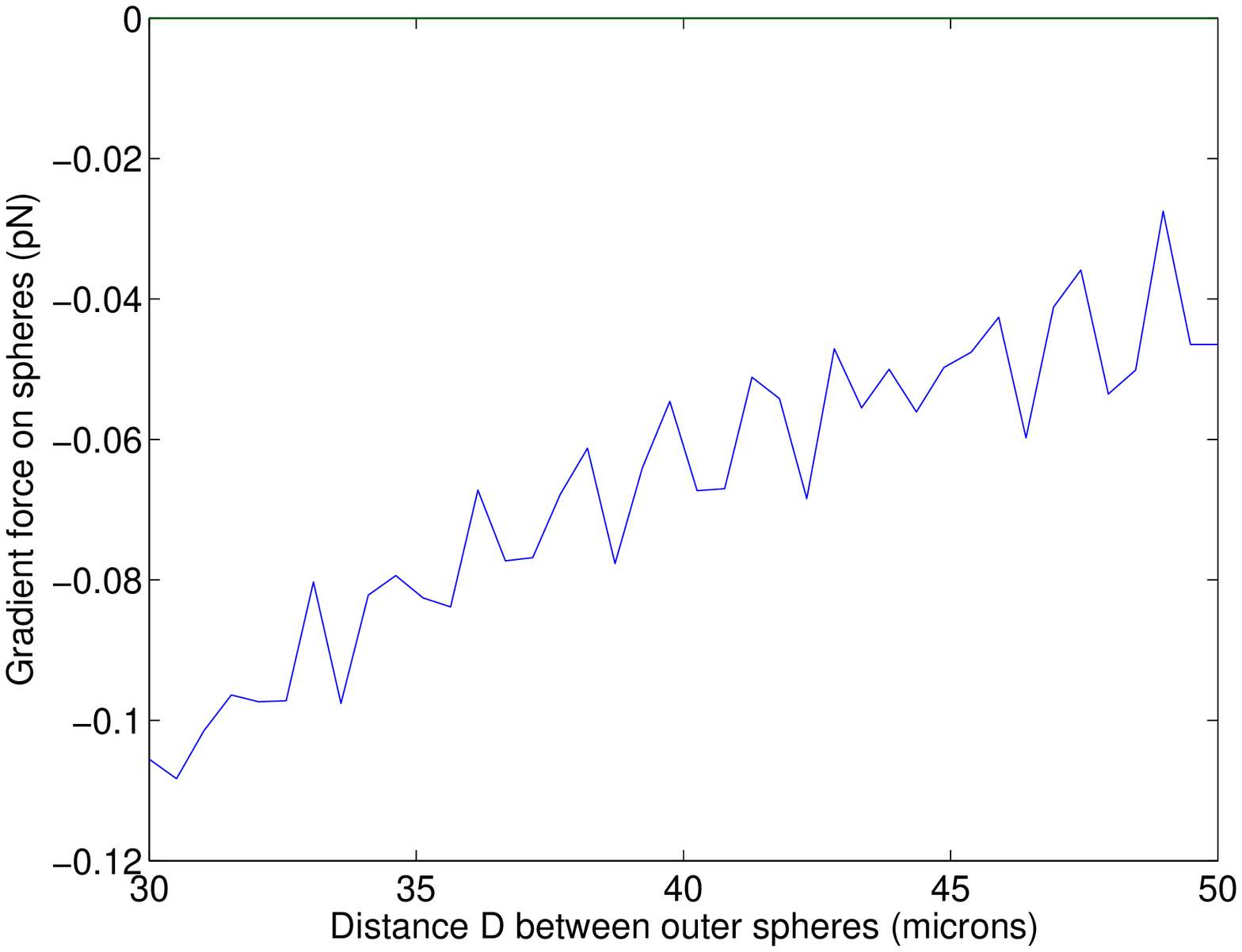}}
\caption{Gradient force on two 2.3 $\mu m$ diameter silica spheres
with the beam waists 180 $\mu m$ apart. $\omega_0=3.5\mu m$ and
$\lambda=1064nm$.}
\end{figure}

\begin{figure}
\centerline{\includegraphics[width=8cm]{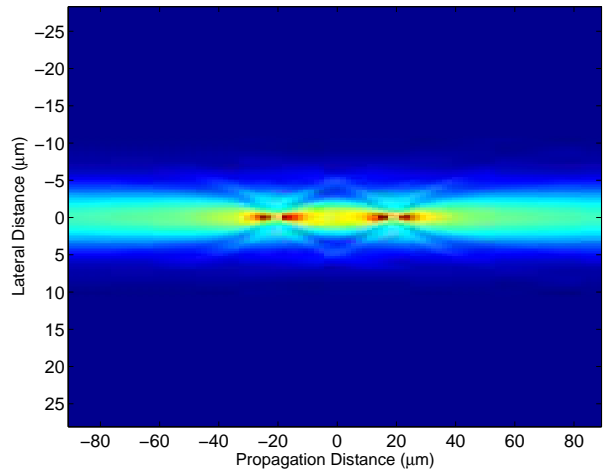}}
\caption{Intensity plot in the x-z plane for the case of two 2.3
$\mu m$ diameter silica spheres with the beam waists 180 $\mu m$
apart. $\omega_0=3.5\mu m$ and $\lambda=1064nm$.}
\end{figure}

\begin{figure}
\centerline{\includegraphics[width=8cm]{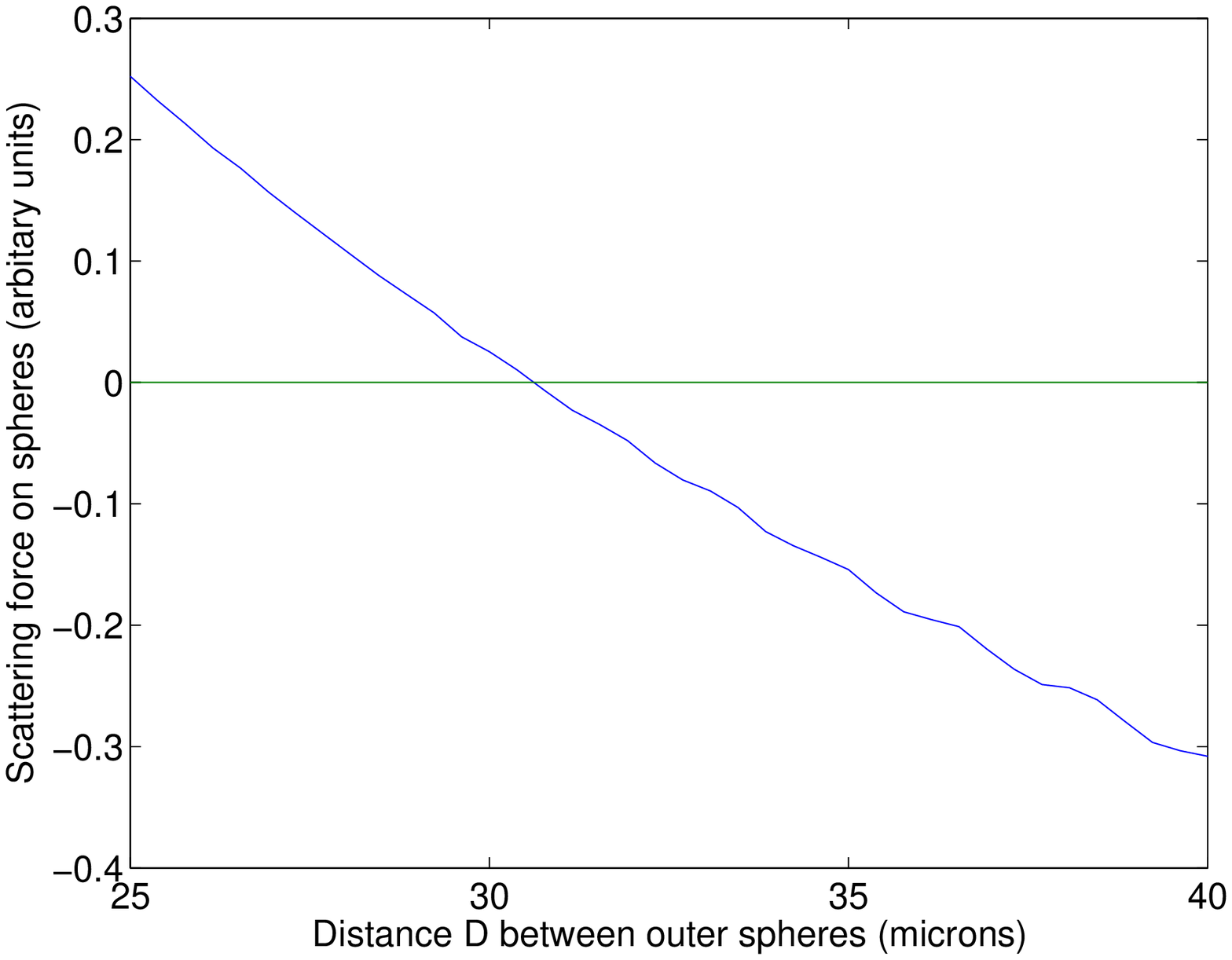}}
\caption{Scattering force on three 2.3 $\mu m$ diameter silica
spheres with the beam waists 180 $\mu m$ apart. $\omega_0=3.5\mu
m$ and $\lambda=1064nm$. The plot shows the separation between two
of the three spheres, and the scattering forces are symmetric
about the center sphere.}
\end{figure}

\subsection{3 micrometer diameter spheres}
The data for 3 micron spheres carried out at a different
wavelength than the 2.3 $\mu m$ data ($\lambda=780nm$) also fits
well with our theory. For two spheres, with the beam waists
$150\mu m$ apart, we predict a sphere separation of ~47$\mu m$
(figure 6) while our experiment predicts a distance of ~45$\mu m$.
Using the same parameters for the three sphere case we predict a
sphere separation of 27$\mu m$ (figure 7), while our experiment
shows a separation of ~35$\mu m$. Again, as we predict equilibrium
positions with the scattering force component, but not with the
gradient force component, we conclude that the scattering force is
the dominant factor in determining the final sphere separations.

\begin{figure}
\centerline{\includegraphics[width=8cm]{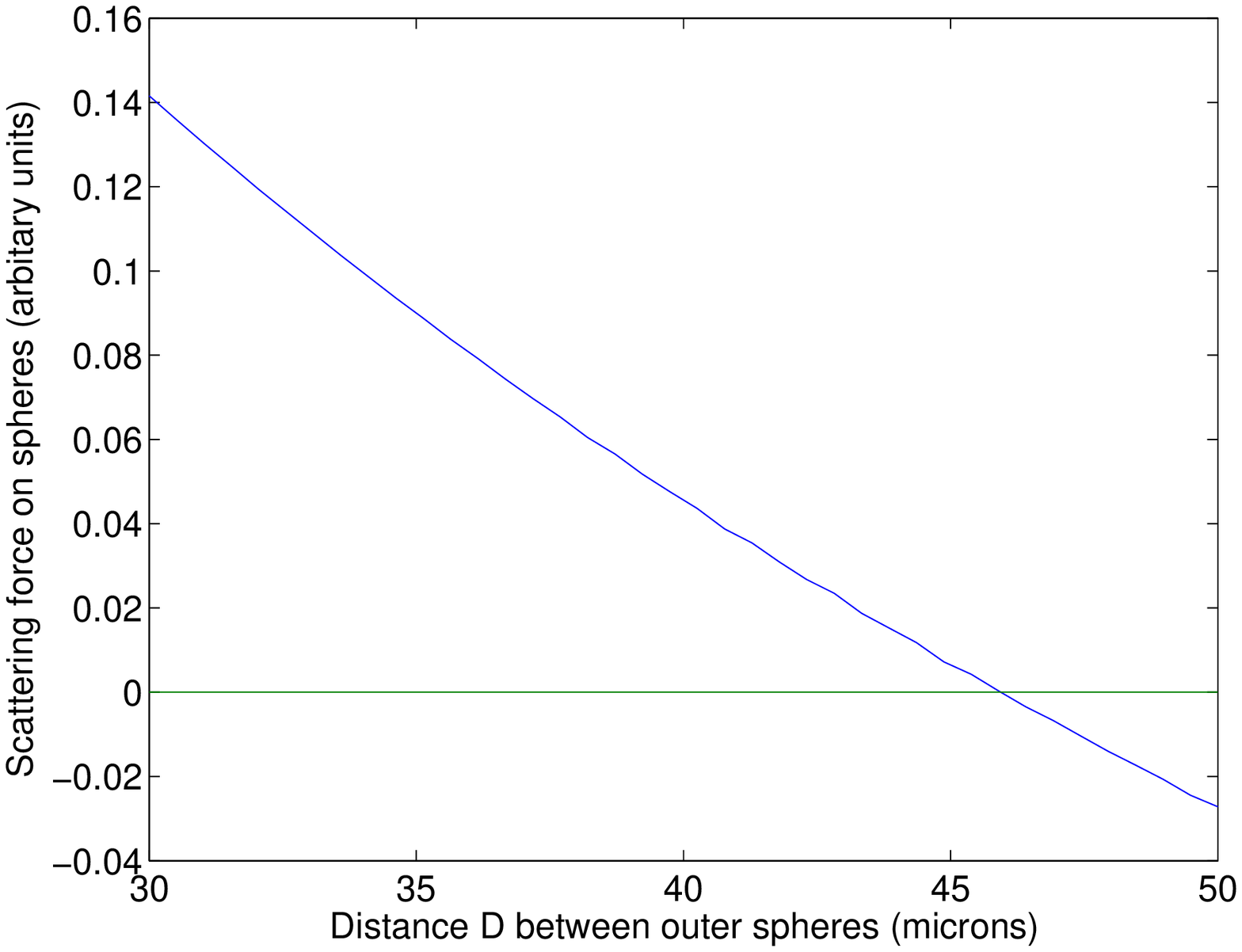}}
\caption{Scattering force on two 3 $\mu m$ diameter silica spheres
with the beam waists 150 $\mu m$ apart. $\omega_0=4.3\mu m$ and
$\lambda=780nm$.}
\end{figure}

\begin{figure}
\centerline{\includegraphics[width=8cm]{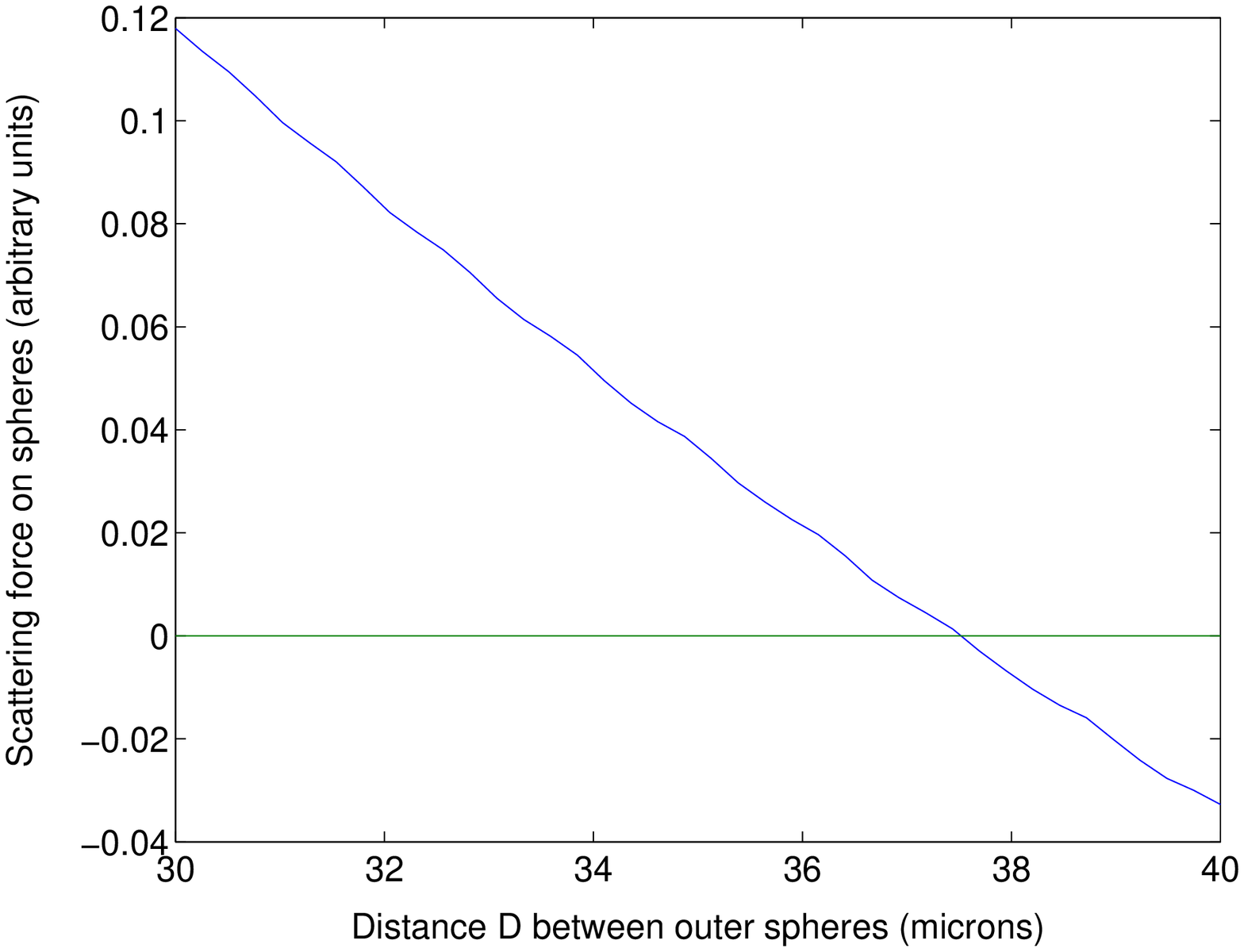}}
\caption{Scattering force on three 3 $\mu m$ diameter silica
spheres with the beam waists 150 $\mu m$ apart. $\omega_0=4.3\mu
m$ and $\lambda=780nm$. The plot shows the separation between two
of the three spheres, and the scattering forces are symmetric
about the center sphere.}
\end{figure}

\section{Discussion and Conclusions}

Our model accurately predicts separations for the case of two and
three spheres, at certain sizes. However we also performed
experiments using $1\mu m$ diameter spheres and could not find any
agreement between experiment and theory. Since our model uses a
paraxial approximation, the assumption is that in these smaller
size regimes the model breaks down. This in contrast to the work
detailed in \cite{Singer} which works in size regimes closer to
the laser wavelength, $\lambda$, and begins to break down in the
larger size regimes ($\lambda > 2l$), where $l$ is the sphere
diameter.

We also note that the beam separation distance becomes less
critical as it becomes larger. For small beam waist separation
distances ($l<<80\mu m$, say), any change in this parameter leads
to a sharp change in the sphere separation distance, whereas at
the waist separation distances we work at the change in sphere
separation distance is far more gentle, and hence gives less rise
to uncertainty over exact fits with theory and experiment. The
other main parameter is sphere size, which has an appreciable
effect on the predicted sphere separation. The incident power on
the spheres does not make much of a difference and is more of a
scaling factor in the forces involved rather than a direct
modifier in the model. Predicted sphere separation is also
sensitive to the refractive index difference between the spheres
and the surrounding medium, so it is important that the spheres'
refractive index is well known.

It should also be possible to create two-dimensional and possibly
three dimensional arrays from optically bound matter. The
extension to two dimensions is relatively simple to envisage with
the use of multiple pairs of counterpropagating laser beams. In
three dimensions the formation of such optically bound arrays may
circumvent some of the problems associated with loading of
three-dimensional optical lattices \cite{vanB}. It is often
assumed that the creation of an optical lattice (via multibeam
interference, say) will allow the simple, unambiguous trapping of
particles in all the lattice sites, thereby making an extended
three-dimensional array of particles. Such arrays may be useful
for crystal template formation \cite{vanB} and in studies of
crystallization processes \cite{Bechinger,Reichhardt}. However
crystal formation in this manner is not particularly robust in
that as the array is filled the particles perturb the propagating
light field such that they prevent the trap sites below them being
efficiently filled. Arrays of optically bound matter do not suffer
from such problems, as they are organized as a result of the
perturbation of the propagating fields. Further the fact that the
particles are bound together provides more realistic opportunities
for studying crystal and colloidal behaviour than that in unbound
optically generated arrays, such as those produced holographically
\cite{Grier1,Grier2,Bechinger}.

We have developed a model by which the propagation of
counter-propagating lasers beams moving past an array of silica
spheres may be examined. Analysis of the resulting forces on the
spheres allows us to predict the separation of the spheres which
constitute the array. We have compared this model with
experimental results for different beam parameters (wavelength,
waist separation, waist diameter) and found the results to be in
good agreement with our observations. The model, does not however,
work with sphere sizes much less than approximately twice the
laser wavelength. Our model is readily extendable to larger number
of spheres, and will be of great use in the study of such one- and
higher-dimensional arrays of optically bound matter.

\section*{Acknowledgements}

DM is a Royal Society University Research Fellow. This work is
supported by the Royal Society and the UK's EPSRC.

\end{document}